\documentclass{paper}

\usepackage{amsmath}
\usepackage{amsthm}
\usepackage{multirow}
\usepackage{longtable}
\usepackage{url}
\usepackage{booktabs}

\begin{document}
\title{A Memetic Algorithm for the Generalized Traveling Salesman Problem
\thanks{This is a modified version of the paper ``A Memetic Algorithm for the Generalized Traveling Salesman Problem'' by G. Gutin, D. Karapetyan and N. Krasnogor published in the proceedings of NICSO 2007.}}

\author{Gregory Gutin\thanks{Department of Computer Science, Royal Holloway University of London, Egham, Surrey TW20 0EX, UK, \texttt{gutin@cs.rhul.ac.uk}} \and Daniel Karapetyan\thanks{Department of Computer Science, Royal Holloway University of London, Egham, Surrey TW20 0EX, UK, \texttt{daniel.karapetyan@gmail.com}}}


\date{}

\maketitle

\begin{abstract} The generalized traveling salesman problem (GTSP) is an extension of the well-known traveling salesman problem.  In GTSP, we are given a partition of cities into groups and we are required to find a minimum length tour that includes exactly one city from each group.  The recent studies on this subject consider different variations of a memetic algorithm approach to the GTSP.  The aim of this paper is to present a new memetic algorithm for GTSP with a powerful local search procedure.  The experiments show that the proposed algorithm clearly outperforms all of the known heuristics with respect to both solution quality and running time.  While the other memetic algorithms were designed only for the symmetric GTSP, our algorithm can solve both symmetric and asymmetric instances.

\end{abstract}

\section{Introduction}

The \emph{generalized traveling salesman problem} (GTSP) is defined as follows. We are given a weighted complete directed or undirected graph $G$ and a partition $V = V_1 \cup V_2 \cup \ldots \cup V_M$ of its vertices; the subsets $V_i$ are called {\em clusters}. The objective is to find a minimum weight cycle containing exactly one vertex from each cluster. There are many publications on GTSP (see, e.g., the surveys~\cite{fischetti2002,gutin2003} and the references there) and the problem has many applications, see, e.g.,~\cite{Transformation,Applications}.  The problem is NP-hard, since the \emph{traveling salesman problem} (TSP) is a special case of GTSP when $|V_i| = 1$ for each $i$.  GTSP is trickier than TSP in the following sense: it is an NP-hard problem to find a minimum weight collection of vertex-disjoint cycles such that each cluster has only one vertex in the collection (and the claim holds even when each cluster has just two vertices)~\cite{gutinAJC27}.  Compare it with the well-known fact that a minimum weight collection of vertex-disjoint cycles in a weighted complete digraph can be found in polynomial time~\cite{TSPAndVariations}.

We call GTSP and TSP \emph{symmetric} if the complete graph $G$ is undirected and \emph{asymmetric} if $G$ is directed. Often instead of the term weight we use the term \emph{length}.

Various approaches to GTSP have been studied.  There are exact algorithms such as branch-and-bound and branch-and-cut algorithms in~\cite{BranchAndCut}. While exact algorithms are very important, they are unreliable with respect to their running time that can easily reach many hours or even days. For example, the well-known TSP solver {\sc Concorde} can easily solve some TSP instances with
several thousand cities, but it could not solve several asymmetric instances with 316 cities within the time limit of $10^4$ sec. (in fact, it appears it would fail even if significantly much more time was allowed)~\cite{BranchAndCut}.

Several researchers use transformations from GTSP to TSP~\cite{Transformation} as there exists a large variety of exact and heuristic algorithms for the TSP, see, e.g.,~\cite{TSPAndVariations,TSPCombOpt}. However, while the known transformations normally allow to produce GTSP optimal solutions from the obtained optimal TSP tours, all known transformations do not preserve suboptimal solutions. Moreover, conversions of near-optimal TSP tours may well result in infeasible GTSP solutions. Thus, the transformation do not allow us to obtain quickly approximate GTSP solutions and there is a necessity for specific GTSP heuristics. Not every TSP heuristic can be extended to GTSP; for example, so-called subtour patching heuristics often used for the Asymmetric TSP, see, e.g.,~\cite{johnson2002a}, cannot be extended to GTSP due to the above mentioned NP-hardness result from \cite{gutinAJC27}.

It appears that the only metaheuristic algorithms that can compete with Lin-Kirnighan-based local search for TSP are memetic algorithms~\cite{KrasnogorBook,moscato1999} that combine powers of genetic and local search algorithms~\cite{johnson2002s,tsai}.  Thus, it is no coincidence that the latest studies in the area of GTSP explore the memetic algorithm approach~\cite{GoldenGA,RandomKeyGA,FatihGA}.

The aim of this paper is to present a new memetic algorithm for GTSP with a powerful local search part.  Unlike the previous heuristics which can be used for the symmetric GTSP only, our algorithm can be used for both symmetric and asymmetric GTSPs.  The computational experiments show that our algorithm clearly outperforms all published memetic heuristics~\cite{GoldenGA,RandomKeyGA,FatihGA} with respect to both solution quality and running time.

\section{The Genetic Algorithm}
\label{sec:ga}

Our heuristic is a memetic algorithm, which combines power of genetic algorithm with that of local search~\cite{KrasnogorBook,Krasnogor2005}.  We start with a general scheme of our heuristic, which is similar to the general schemes of many memetic algorithms.

\begin{itemize}
	\item[Step 1] \emph{Initialize}.  Construct the first generation of solutions.  To produce a solution we use a semirandom construction heuristic (see Subsection~\ref{subsec:first_generation}).

	\item[Step 2] \emph{Improve}.  Use a local search procedure to replace each of the first generation solutions by the local optimum.  Eliminate duplicate solutions.

	\item[Step 3] \emph{Produce next generation}.  Use reproduction, crossover, and mutation genetic operators to produce the non-optimized next generation.  Each of the genetic operators selects parent solutions from the previous generation.  The length of a solution is used as the evaluation function.

	\item[Step 4] \emph{Improve next generation}.  Use a local search procedure to replace each of the current generation solutions except the reproduced ones by the local optimum.  Eliminate duplicate solutions.

	\item[Step 5] \emph{Evolute}.  Repeat Steps 3--4 until a termination condition is reached.
\end{itemize}

\subsection{Coding}

The Genetic Algorithm (GA) requires each solution to be coded in a \emph{chromosome}, i.e., to be represented by a sequence of \emph{genes}.  Unlike~\cite{RandomKeyGA,FatihGA} we use a natural coding of the solutions as in~\cite{GoldenGA}.  The coded solution is a sequence of numbers ($s_1$ $s_2$ \ldots $s_M$) such that $s_i$ is the vertex at the position $i$ of the solution.  For example (2~5~9~4) represents the cycle visiting vertex 2, then vertex 5, then vertex 9, then vertex 4, and then returning to vertex 2.  Note that not any sequence corresponds to a feasible solution as the feasible solution should contain exactly one vertex from each cluster, i.e., $C(s_i) \neq C(s_j)$ for any $i \neq j$, where $C(v)$ is the cluster containing vertex $v$.

Note that, using natural coding, each solution can be represented by $M$ different chromosomes: the sequence can be `rotated', i.e., the first gene can be moved to the end of the chromosome or the last gene can be inserted before the first one and these operations will preserve the cycle.  For example, chromosomes (2~5~9~4) and (5~9~4~2) represent the same solution.  We need to take this into account when considering several solutions together, i.e., in precisely two cases: when we compare two solutions, and when we apply crossover operator.  In these cases we `normalise' the chromosomes by rotating each of them such that the vertex $v \in V_1$ (the vertex that represents the cluster 1) takes the first place in the chromosome.  For example, if we had a chromosome (2~5~9~4) and the vertex 5 belongs to the cluster 1, we rotate the chromosome in the following way: (5~9~4~2).

In the case of the symmetric problem the chromosome can also be `reflected' while preserving the solution.  But our heuristic is designed for both symmetric and asymmetric instances and, thus, the chromosomes (1~5~9~4) and (4~9~5~1) are considered as the chromosomes corresponding to distinct solutions.

The main advantage of the natural coding is its efficiency in the local search.  As the local search is the most time consuming part of our heuristic, the coding should be optimized for it.

\subsection{First Generation}
\label{subsec:first_generation}

We produce $2 M$ solutions for the first generation, where $M$ is the number of clusters.  The solutions are generated by a \emph{semirandom construction heuristic}.  The semirandom construction heuristic generates a random cluster permutation and then finds the best vertex in each cluster when the order of clusters is given by the permutation.

It chooses the best vertex selection within the given cluster sequence using the Cluster Optimization Heuristic (see Section~\ref{sec:local_improvement}).

The advantages of the semirandom construction heuristic are that it is fast and its cycles have no regularity.  The latter is important as each completely deterministic heuristic can cause solutions uniformity and as a result some solution branches can be lost.

\subsection{Next Generations}
\label{sec:next_generations}

Each generation except the first one is based on the previous generation.  To produce the next generation one uses genetic operators, which are algorithms that construct a solution or two from one or two so-called parent solutions.  Parent solutions are chosen from the previous generation using some \emph{selection strategy}.  We perform $r$ runs of \emph{reproduction}, $8 r$ runs of \emph{crossover}, and $2 r$ runs of \emph{mutation} operator.  The value $r$ is calculated as $r = 0.2 G + 0.05 M + 10$, where $G$ is the number of generations produced before the current one.  (Recall that $M$ is the number of clusters.)  As a result, we obtain at most $11 r$ solutions in each generation but the first one (since we remove duplicated solutions from the population, the number of solutions in each generation can be smaller than $11 r$).  From generation to generation, one can expect the number of local minima found by the algorithm to increase.  Also this number can be expected to grow when the number of clusters $M$ grows.  Thus, in the formula above $r$ depends on both $G$ and $M$.  All the coefficients in the formulas of this section were obtained in computational experiments, where several other values of the coefficients were also tried.  Note that slight variations in selection of the coefficients do not influence significantly the results of the algorithm.

\subsection{Reproduction}
\label{sec:reproduction}

Reproduction is a process of simply copying solutions from the previous generation.  Reproduction operator requires a selection strategy to select the solutions from the previous generation to be copied.  In our algorithm we select $r$ (see Subsection~\ref{sec:next_generations}) shortest solutions from the previous generation to copy them to the current generation.

\subsection{Crossover}

A \emph{crossover} operator is a genetic operator that combines two different solutions from the previous generation.  We use a modification of the two-point crossover introduced by Silberholz and Golden~\cite{GoldenGA} as an extension of an \emph{Ordered Crossover}~\cite{Davis}.  Our crossover operator produces just one child solution $(r_1~r_2~\ldots~r_M)$ from the parent solutions $(p_1~p_2~\ldots~p_M)$ and $(q_1~q_2~\ldots~q_M)$.  At first it selects a random position $a$ and a random fragment length $1 \le l < M$ and copies the fragment $[a, a + l)$ of the first parent to the beginning of the child solution: $r_i = p_{i+a}$ for each $i = 0, 1, \ldots, l - 1$.\footnote{We assume that $s_{i+M} = s_i$ for the solution $(s_1~s_2~\ldots~s_M)$ and for any $1 \le i \le M$.}  To produce the rest of the child solution, we introduce a sequence $q'$ as follows: $q'_i = q_{i+a+l-1}$, where $i = 1, 2, \ldots, M$.  Then for each $i$ such that the cluster $C(q'_i)$ is already visited by the child solution $r$, the vertex $q'_i$ is removed from the sequence: $q' = (q'_1~q'_2\ldots~q'_{i-1}~q'_{i+1}\ldots)$.  As a result $l$ vertices will be removed: $|q'| = M - l$.  Now the child solution $r$ should be extended by the sequence $q'$: $r = (r_1~r_2~\ldots~r_l~q'_1~q'_2~\ldots~q'_{M-l})$.

A feature of this crossover is that it preserves the vertex order of both parents.

\emph{Crossover example}. Let the first parent be $(1~2~3~4~5~6~7)$ and the second parent $(3~2~5~7~6~1~4)$ (here we assume for explanation clarity that every cluster contains exactly one vertex: $V_i = \{ i \}$).  First of all we rotate the parent solutions such that $C(p_1) = C(q_1) = 1$: $p = (1~2~3~4~5~6~7)$ (remains the same) and $q = (1~4~3~2~5~7~6)$.  Now we choose a random fragment in the parent solutions:\\
$p = (1~2~|~3~4~|~5~6~7)$\\
$q = (1~4~|~3~2~|~5~7~6)$\\
and copy this fragment from the first parent $p$ to the child solution: $r = (3~4)$.  Next we produce the sequence $q' = (5~7~6~1~4~3~2)$ and remove vertices 3 and 4 from it as the corresponding clusters are already visited by $r$: $q' = (5~7~6~1~2)$.  Finally, we extend the child solution $r$ by $q'$:\\
$r = (3~4~5~7~6~1~2)$.
\quad

The crossover operator requires some strategy to select two parent solutions from the previous generation.  In our algorithm an elitist strategy is used; the parents are chosen randomly between the best $33 \%$ of all the solutions in the previous generation.

\subsection{Mutation}

A \emph{mutation} operator modifies partially some solution from the previous generation.  The modification should be stochastic and usually worsens the solution.  The goal of the mutation is to increase the solution diversity in the generation.

Our mutation operator removes a random fragment of the solution and inserts it in some random position.  The size of the fragment is selected between $0.05 M$ and $0.3 M$.  An elitist strategy is used in our algorithm; the parent is selected randomly among $75 \%$ of all the solutions in the previous generation.

\emph{Mutation example}. Let the parent solution be (1~2~3~4~5~6~7).  Let the random fragment start at 2 and be of the length 3.  The new fragment position is 3, for example.  After removing the fragment we have (1~5~6~7).  Now insert the fragment (2~3~4) at the position 3: (1~5~2~3~4~6~7).

\subsection{Termination condition}

For the termination condition we use the concept of idle generations.  We call a generation \emph{idle} if the best solution in this generation has the same length as the length of the best solution in the previous generation.  In other words, if the produced generation has not improved the solution, it is idle.  The heuristic stops after some idle generations are produced sequentially.

In particular, we implemented the following new condition.  Let $I(l)$ be the number of sequential idle generations with the best solution of length $l$.  Let $I_{cur} = I(l_{cur})$, where $l_{cur}$ is the current best solution length.  Let $I_{max} = \max_{l > l_{cur}} I(l)$.  Then our heuristic stops if $I_{cur} \ge \max(1.5 I_{max}, 0.05 M + 5)$.  This formula means that we are ready to wait for the next improvement 1.5 times more generations than we have ever waited previously.  The constant $0.05 M + 5$ is the minimum boundary for the number of generations we are ready to wait for improvement.  All the coefficients used in the formula were found empirically.

\subsection{Asymmetric instances}

Our algorithm is designed to process equally both symmetric and asymmetric instances, however some parameters should take different values for these types of instances for the purpose of high efficiency.  In particular, we double the size of the first generation ($4 M$ instead of $2 M$, see Subsection~\ref{subsec:first_generation}) and increase the minimum number of idle generations by 5 (i.e., $I_{cur} \ge \max(1.5 I_{max}, 0.05 M + 10)$).  The local improvement procedure (see below) has also some differences for symmetric and asymmetric instances.

\section{Local Improvement Part}
\label{sec:local_improvement}

We use a \emph{local improvement procedure} for each solution added to the current generation.  The local improvement procedure runs several local search heuristics sequentially.  The following local search heuristics are used in our algorithm:

\begin{itemize}
	\item \emph{Swaps} tries to swap every non-neighboring pair of vertices.  The heuristic applies all the improvements found during one cycle of swaps.

	\item \emph{$k$-Neighbor Swap} tries different permutations of every solution subsequence $(s_1~s_2~\ldots~s_k)$.  In particular it tries all the non-trivial permutations which are not covered by any of $i$-Neighbor Swap, $i = 2, 3, \ldots, k-1$.  For each permutation the best selection of the vertices within the considered cluster subsequence is calculated.  The best permutation is accepted if it improves the solution.  The heuristic applies all the improvements found during one cycle.



	\item \emph{2-opt} tries to replace every non-adjacent pair of edges $s_i s_{i+1}$ and $s_j s_{j+1}$ in the solution by the edges $s_i s_j$ and $s_{i+1} s_{j+1}$ if the new edges are lighter, i.e., the sum of their weights is smaller than the sum of the weights of old edges.  The heuristic applies all the improvements found.

	\item \emph{Direct 2-opt} is a modification of 2-opt heuristic.  Direct 2-opt selects a number of the longest edges contained in the solution and then tries all the non-adjacent pairs of the selected edges.  It replaces edges $s_i s_{i+1}$ and $s_j s_{j+1}$ with the edges $s_i s_j$ and $s_{i+1} s_{j+1}$ if the new edges are shorter, i.e., the sum of their weights is smaller than the sum of the weights of old edges.  The heuristic applies all the improvements found.

	\item \emph{Inserts} tries to remove a vertex from the solution and to insert it in the different position.  The best vertex in the inserted cluster is selected after the insertion.  The insertion is accepted if it improves the solution.  The heuristic tries every combination of the old and the new positions except the neighboring positions and applies all the improvements found.


	\item \label{item:cluster_optimisation} \emph{Cluster Optimization} (CO) uses the shortest $(s, t)$-path algorithm for acyclic digraphs (see, e.g.,~\cite{bang2000}) to find the best vertex for each cluster when the order of clusters is fixed.  This heuristic was introduced by Fischetti, Salazar-Gonz{\'a}lez and Toth~\cite{BranchAndCut} (see its detailed description also in~\cite{fischetti2002}).

The CO Heuristic uses the fact that the shortest $(s, t)$-path in an acyclic digraph can be found in a polynomial time.  Let the given solution be represented by chromosome $(s_1~s_2~\ldots~s_M)$.  The algorithm builds an acyclic digraph $G_\text{CO} = (V_\text{CO}, E_\text{CO})$, where $V_\text{CO} = V \cup C'(s_1)$ is the set of the GTSP instance vertices extended by a copy of the cluster $C(s_1)$ and $E_\text{CO}$ is a set of edges in the digraph $G_\text{CO}$.  (Recall that $C(x)$ is the cluster containing the vertex $x$.)  An edge $xy \in E_\text{CO}$ if and only if $C(x) = C(s_i)$ and $C(y) = C(s_{i+1})$ for some $i < M$ or if $C(x) = C(s_M)$ and $C(y) = C'(s_{1})$.  For each vertex $s \in C(s_1)$ and its copy $s' \in C'(s_1)$, the algorithm finds the shortest $(s, s')$-path in $G_\text{CO}$.  The algorithm selects the shortest path $(s~p_2~p_3~\ldots~p_M~s')$ and returns the chromosome $(s~p_2~p_3~\ldots~p_M)$ which is the best vertex selection within the given cluster sequence.

Note that the algorithm's time complexity grows linearly with the size of the cluster $C(s_1)$.  Thus, before applying the CO algorithm we rotate the initial chromosome in such a way that $|C(s_1)| = \min_{i \le M} |C_i|$.
\end{itemize}

For each local search algorithm with some cluster optimization embedded, i.e., for $k$-Neightbour Swap and Inserts, we use a speed-up heuristic.  We calculate a lower bound $l_\text{new}$ of the new solution length and compare it with the previous length $l_\text{prev}$ before the vertices within the clusters optimization.  If $l_\text{new} \ge l_\text{prev}$, the solution modification is declined immediately.  For the purpose of the new length lower bound calculation we assume that the unknown edges, i.e., the edges adjacent to the vertices that should be optimized, have the length of the shortest edges between the corresponding clusters.

Some of these heuristics form a heuristic-vector $\mathcal{H}$ as follows:

\begin{tabular}{ll}
Symmetric instances								& Asymmetric instances \\
\cmidrule(){1-2}
Inserts												& Swaps \\
Direct 2-opt for $M / 4$ longest edges		& Inserts \\
2-opt								& Direct 2-opt for $M / 4$ longest edges \\
2-Neighbour Swap				& 2-opt \\
3-Neighbour Swap				& 2-Neighbour Swap \\
4-Neighbour Swap				& 3-Neighbour Swap \\
\cmidrule(){1-2}
\end{tabular}

The improvement procedure applies all the local search heuristic from $\mathcal{H}$ cyclically.  Once some heuristic fails to improve the tour, it is excluded from $\mathcal{H}$.  If 2-opt heuristic fails, we also exclude Direct 2-opt from $\mathcal{H}$.  Once $\mathcal{H}$ is empty, the CO heuristic is applied to the solution and the improvement procedure stops.

\section{Results of Computational Experiments}

We tested our heuristic using GTSP instances which were generated from some TSPLIB~\cite{TSPLIB} instances by applying the standard clustering procedure of Fischetti, Salazar, and Toth~\cite{BranchAndCut}.  Note that our heuristic is designed for medium and large instances and, thus, we selected all the instances with 40 to 217 clusters.  Unlike~\cite{GoldenGA,RandomKeyGA,FatihGA}, smaller instances are not considered.

All the information necessary for reproducing our experiments is available online at \url{www.cs.rhul.ac.uk/Research/ToC/publications/Karapetyan}:
\begin{itemize}
	\item All the instances considered in our experiments.  For the purpose of simplicity and efficiency we use a uniform binary format for instances of all types.

	\item The binary format definition.

	\item Source codes of binary format reading and writing procedures.

	\item Source codes of the clustering procedure~\cite{BranchAndCut} to convert TSP instances into GTSP instances.

	\item Source codes of the TSPLIB files reading procedure.

	\item Source codes of our memetic algorithm.

	\item Source codes of our experimentation engine.
\end{itemize}

The tables below show the experiments results.  We compare the following heuristics:
\begin{description}
	\item[GK] is the heuristic presented in this paper.
	\item[SG] is the heuristic by Silberholz and Golden~\cite{GoldenGA}.
	\item[SD] is the heuristic by Snyder and Daskin~\cite{RandomKeyGA}.
	\item[TSP] is the heuristic by Tasgetiren, Suganthan, and Pan~\cite{FatihGA}.
\end{description}

The results for \texttt{GK} and \texttt{SD} were obtained in our own experiments.  Other results are taken from the corresponding papers.  Each test of \texttt{GK} and \texttt{SD} includes ten algorithm runs.  The results for \texttt{SG} and \texttt{TSP} were produced after five runs.

To compare the running times of all the considered heuristics we need to convert the running times of \texttt{SG} and \texttt{TSP} obtained from the corresponding papers to the running times on our evaluation platform.  Let us assume that the running time of some Java implemented algorithm on the \texttt{SG} evaluation platform is $t_\text{SG} = k_\text{SG} \cdot t_\text{GK}$, where $k_\text{SG}$ is some constant and $t_\text{GK}$ is the running time of the same but \texttt{C++} implemented algorithm on our evaluation platform.  Let us assume that the running time of some algorithm on the \texttt{TSP} evaluation platform is $t_\text{TSP} = k_\text{TSP} \cdot t_\text{GK}$, where $k_\text{TSP}$ is some constant and $t_\text{GK}$ is the running time of the same algorithm on our evaluation platform.

The computer used for \texttt{GK} and \texttt{SD} evaluation has the AMD Athlon 64 X2 3.0~GHz processor.  The computer used for \texttt{SG} has Intel Pentium 4 3.0~GHz processor.  The computer used for \texttt{TSP} has Intel Centrino Duo 1.83~GHz processor.  Heuristics \texttt{GK}, \texttt{SD}, and \texttt{TSP} are implemented in \texttt{C++} (\texttt{GK} is implemented in \texttt{C\#} but the most time critical fragments are implemented in \texttt{C++}).  Heuristic \texttt{SG} is implemented in \texttt{Java}.  Some rough estimation of \texttt{Java} performance in the combinatorial optimisation applications shows that \texttt{C++} implementation could be approximately two times faster than the \texttt{Java} implementation.  As a result the adjusting coefficient $k_\text{SG} \approx 3$ and the adjusting coefficient $k_\text{TSP} \approx 2$.

We are able to compare the results of \texttt{SD} heuristic tests gathered from different papers to check the $k_{SG}$ and $k_{TSP}$ values because \texttt{SD} has been evaluated on each of the platforms of our interest (the heuristic was implemented in \texttt{Java} in~\cite{GoldenGA} for the exact comparison to \texttt{SG}).  The time ratio between the \texttt{SD} running times from~\cite{GoldenGA} and our own results vary significantly for different problems, but for some middle size problems the ratio is about 2.5 to 3.  These results correlate well with the previous estimation.  The suggested value $k_\text{TSP} \approx 2$ is also confirmed by this method.

\bigskip

The headers of the tables in this section are as follows:
\begin{description}
	\item[Name] is the instance name.  The prefix number is the number of clusters in the instance; the suffix number is the number of vertices.

	\item[Error, \%] is the error, in per cent, of the average solution above the optimal value.  The error is calculated as $\frac{value - opt}{opt} \times 100 \%$, where $value$ is the obtained solution length and $opt$ is the optimal solution length.  The exact optimal solutions are known from~\cite{Transformation} and from~\cite{BranchAndCut} for 17 of the considered instances only.  For the rest of the problems we use the best solutions ever obtained in our experiments instead.

	\item[Time, sec] is the average running time for the considered heuristic in seconds.  The running times for \texttt{SG} and for \texttt{TSP} are obtained from the corresponding papers thus these values should be adjusted using $k_\text{SG}$ and $k_\text{TSP}$ coefficients, respectively, before the comparison.

	\item[Quality impr., \%] is the improvement of the average solution quality of the \texttt{GK} with respect to some other heuristic.  The improvement is calculated as $E_H - E_\text{GK}$ where $E_H$ is the average error of the considered heuristic $H$ and $E_\text{GK}$ is the average error of our heuristic.

	\item[Time impr.] is the improvement of the \texttt{GK} average running time with respect to some other heuristic running time.  The improvement is calculated as $T_H / T_\text{GK}$ where $T_H$ is the average running time of the considered heuristic $H$ and $T_\text{GK}$ is the average running time of our heuristic.

	\item[Opt., \%] is the number of tests, in per cent, in which the optimal solution was reached.  The value is displayed for three heuristics only as we do not have it for \texttt{SG}.

	\item[Opt.] is the best known solution length.  The exact optimal solutions are known from~\cite{BranchAndCut} and~\cite{Transformation} for 17 of the considered instances only.  For the rest of the problems we use the best solutions ever obtained in our experiments.

	\item[Value] is the average solution length.

	\item[\# gen.] is the average number of generations produced by the heuristic.
\end{description}

\begin{table}
	\caption{Solvers quality comparison.}
	\label{tab:quality}
\begin{tabular}{l|rrrr|rrr|rrr}

\multirow{2}{*}{Name}		&	\multicolumn{4}{c|}{Error, \%}							&	\multicolumn{3}{c|}{Quality impr., \%}					&	\multicolumn{3}{c}{Opt., \%}					 \\
		&	GK	&	SG	&	SD	&	TSP	&	SG	&	SD	&	TSP	&	GK	&	SD	&	TSP	\\
\hline																						
40d198		&	0.00	&	0.00	&	0.00	&	0.00	&	0.00	&	0.00	&	0.00	&	100	&	100	&	100	\\
40kroa200		&	0.00	&	0.00	&	0.00	&	0.00	&	0.00	&	0.00	&	0.00	&	100	&	100	&	100	\\
40krob200		&	0.00	&	0.05	&	0.01	&	0.00	&	0.05	&	0.01	&	0.00	&	100	&	70	&	100	\\
41gr202		&	0.00	&		&	0.00	&		&		&	0.00	&		&	100	&	100	&		\\
45ts225		&	0.00	&	0.14	&	0.09	&	0.04	&	0.14	&	0.09	&	0.04	&	100	&	0	&	60	\\
45tsp225		&	0.00	&		&	0.01	&		&		&	0.01	&		&	100	&	90	&		\\
46pr226		&	0.00	&	0.00	&	0.00	&	0.00	&	0.00	&	0.00	&	0.00	&	100	&	100	&	100	\\
46gr229		&	0.00	&		&	0.03	&		&		&	0.03	&		&	100	&	60	&		\\
53gil262		&	0.00	&	0.45	&	0.31	&	0.32	&	0.45	&	0.31	&	0.32	&	100	&	30	&	60	\\
53pr264		&	0.00	&	0.00	&	0.00	&	0.00	&	0.00	&	0.00	&	0.00	&	100	&	100	&	100	\\
56a280		&	0.00	&	0.17	&	0.08	&		&	0.17	&	0.08	&		&	100	&	70	&		\\
60pr299		&	0.00	&	0.05	&	0.05	&	0.03	&	0.05	&	0.05	&	0.03	&	100	&	20	&	60	\\
64lin318		&	0.00	&	0.00	&	0.38	&	0.46	&	0.00	&	0.38	&	0.46	&	100	&	50	&	60	\\
65rbg323	(asym.)	&	0.00	&		&		&		&		&		&		&	100	&		&		\\
72rbg358	(asym.)	&	0.00	&		&		&		&		&		&		&	100	&		&		\\
80rd400		&	0.00	&	0.58	&	0.60	&	0.91	&	0.58	&	0.60	&	0.91	&	100	&	0	&	20	\\
81rbg403	(asym.)	&	0.00	&		&		&		&		&		&		&	100	&		&		\\
84fl417		&	0.00	&	0.04	&	0.02	&	0.00	&	0.04	&	0.02	&	0.00	&	100	&	40	&	100	\\
87gr431		&	0.00	&		&	0.30	&		&		&	0.30	&		&	100	&	40	&		\\
88pr439		&	0.00	&	0.00	&	0.28	&	0.00	&	0.00	&	0.28	&	0.00	&	100	&	20	&	80	\\
89pcb442		&	0.00	&	0.01	&	1.30	&	0.86	&	0.01	&	1.30	&	0.86	&	100	&	0	&	0	\\
89rbg443	(asym.)	&	0.13	&		&		&		&		&		&		&	50	&		&		\\
99d493		&	0.11	&	0.47	&	1.28	&		&	0.36	&	1.17	&		&	10	&	0	&		\\
107ali535		&	0.00	&		&	1.36	&		&		&	1.36	&		&	100	&	0	&		\\
107att532		&	0.01	&	0.35	&	0.72	&		&	0.34	&	0.72	&		&	80	&	0	&		\\
107si535		&	0.00	&	0.08	&	0.32	&		&	0.08	&	0.32	&		&	100	&	0	&		\\
113pa561		&	0.00	&	1.50	&	3.57	&		&	1.50	&	3.57	&		&	100	&	0	&		\\
115u574		&	0.02	&		&	1.54	&		&		&	1.52	&		&	80	&	0	&		\\
115rat575		&	0.20	&	1.12	&	3.22	&		&	0.93	&	3.03	&		&	90	&	0	&		\\
131p654		&	0.00	&	0.29	&	0.08	&		&	0.29	&	0.08	&		&	100	&	0	&		\\
132d657		&	0.15	&	0.45	&	2.32	&		&	0.29	&	2.16	&		&	30	&	0	&		\\
134gr666		&	0.11	&		&	3.74	&		&		&	3.62	&		&	70	&	0	&		\\
145u724		&	0.14	&	0.57	&	3.49	&		&	0.43	&	3.35	&		&	50	&	0	&		\\
157rat783		&	0.11	&	1.17	&	3.84	&		&	1.06	&	3.72	&		&	20	&	0	&		\\
200dsj1000		&	0.12	&		&	2.45	&		&		&	2.33	&		&	30	&	0	&		\\
201pr1002		&	0.14	&	0.24	&	3.43	&		&	0.10	&	3.29	&		&	30	&	0	&		\\
207si1032		&	0.03	&	0.37	&	0.93	&		&	0.34	&	0.91	&		&	20	&	0	&		\\
212u1060		&	0.27	&	2.25	&	3.60	&		&	1.98	&	3.33	&		&	30	&	0	&		\\
217vm1084		&	0.19	&	0.90	&	3.68	&		&	0.71	&	3.49	&		&	60	&	0	&		\\
\hline																						
Full IS average		&	0.04	&		&		&		&		&		&		&	81	&		&		\\
Sym. IS average		&	0.05	&		&	1.43	&		&		&	1.38	&		&	77	&	16	&		\\
SG IS average		&	0.06	&	0.54	&	1.57	&		&	0.47	&	1.50	&		&	72	&	11	&		\\
TSP IS average		&	0.00	&	0.21	&	0.45	&	0.44	&	0.21	&	0.45	&	0.44	&	100	&	17	&	43	\\

\end{tabular}
\end{table}

The results of the experiments presented in Table~\ref{tab:quality} show that our heuristic (\texttt{GK}) has clearly outperformed all other heuristics with respect to solution quality.  For each of the considered instances the average solution reached by our heuristic is always not worse than the average solution reached by any other heuristic and the percent of the runs in which the optimal solution was reached is not less than for any other considered heuristic (note that we are not able to compare our heuristic with \texttt{SG} with respect to this value).

The average values are calculated for four \emph{instance sets} (IS).  The \emph{Full IS} includes all the instances considered in this paper, both symmetric and asymmetric.  The \emph{Sym.~IS} includes all the symmetric instances considered in this paper.  The \emph{SG~IS} includes all the instances considered in both this paper and~\cite{GoldenGA}.  The \emph{TSP~IS} includes all the instances considered in both this paper and~\cite{FatihGA}.

One can see that the average quality of our \texttt{GK} heuristic is approximately 10 times better than that of \texttt{SG} heuristic, approximately 30 times better than that of \texttt{SD}, and for \texttt{TSP} IS our heuristic reaches the optimal solution each run and for each instance, in contrast to \texttt{TSP} that has 0.44\% average error.  The maximum error of \texttt{GK} is 0.27\% while the maximum error of \texttt{SG} is 2.25\% and the maximum error of \texttt{SD} is 3.84\%.

\begin{table}
	\caption{Solvers running time comparison.}
	\label{tab:times}
\begin{tabular}{ l|rrrr|rrr }

\multirow{2}{*}{Name}		&	\multicolumn{4}{c|}{Time, sec}							&	\multicolumn{3}{c}{Time impr., \%}					\\
		&	GK	&	SG	&	SD	&	TSP	&	SG	&	SD	&	TSP	\\
\hline																
40d198		&	0.14	&	1.63	&	1.18	&	1.22	&	11.6	&	8.4	&	8.7	\\
40kroa200		&	0.14	&	1.66	&	0.26	&	0.79	&	12.1	&	1.9	&	5.8	\\
40krob200		&	0.16	&	1.63	&	0.80	&	2.70	&	10.2	&	5.0	&	16.8	\\
41gr202		&	0.21	&		&	0.65	&		&		&	3.2	&		\\
45ts225		&	0.24	&	1.71	&	0.46	&	1.42	&	7.0	&	1.9	&	5.8	\\
45tsp225		&	0.19	&		&	0.55	&		&		&	2.9	&		\\
46pr226		&	0.10	&	1.54	&	0.63	&	0.46	&	15.5	&	6.4	&	4.6	\\
46gr229		&	0.25	&		&	1.14	&		&		&	4.6	&		\\
53gil262		&	0.31	&	3.64	&	0.85	&	4.51	&	11.7	&	2.7	&	14.5	\\
53pr264		&	0.24	&	2.36	&	0.82	&	1.10	&	10.0	&	3.5	&	4.7	\\
56a280		&	0.38	&	2.92	&	1.14	&		&	7.7	&	3.0	&		\\
60pr299		&	0.42	&	4.59	&	1.74	&	3.08	&	10.9	&	4.1	&	7.3	\\
64lin318		&	0.45	&	8.08	&	1.42	&	8.49	&	18.1	&	3.2	&	19.0	\\
65rbg323	(asym.)	&	1.14	&		&		&		&		&		&		\\
72rbg358	(asym.)	&	1.26	&		&		&		&		&		&		\\
80rd400		&	1.07	&	14.58	&	3.53	&	13.55	&	13.7	&	3.3	&	12.7	\\
81rbg403	(asym.)	&	0.98	&		&		&		&		&		&		\\
84fl417		&	0.73	&	8.15	&	3.17	&	6.74	&	11.1	&	4.3	&	9.2	\\
87gr431		&	2.01	&		&	4.01	&		&		&	2.0	&		\\
88pr439		&	1.48	&	19.06	&	4.68	&	20.87	&	12.9	&	3.2	&	14.1	\\
89pcb442		&	1.72	&	23.43	&	4.26	&	23.14	&	13.6	&	2.5	&	13.4	\\
89rbg443	(asym.)	&	3.69	&		&		&		&		&		&		\\
99d493		&	4.17	&	35.72	&	6.34	&		&	8.6	&	1.5	&		\\
107ali535		&	5.82	&		&	7.75	&		&		&	1.3	&		\\
107att532		&	3.45	&	31.70	&	8.04	&		&	9.2	&	2.3	&		\\
107si535		&	1.88	&	26.35	&	6.06	&		&	14.1	&	3.2	&		\\
113pa561		&	3.22	&	21.08	&	6.37	&		&	6.5	&	2.0	&		\\
115u574		&	3.76	&		&	11.48	&		&		&	3.1	&		\\
115rat575		&	4.12	&	48.48	&	9.19	&		&	11.8	&	2.2	&		\\
131p654		&	2.82	&	32.67	&	13.23	&		&	11.6	&	4.7	&		\\
132d657		&	6.82	&	132.24	&	15.40	&		&	19.4	&	2.3	&		\\
134gr666		&	14.46	&		&	21.06	&		&		&	1.5	&		\\
145u724		&	11.61	&	161.82	&	22.00	&		&	13.9	&	1.9	&		\\
157rat783		&	15.30	&	152.15	&	22.70	&		&	9.9	&	1.5	&		\\
200dsj1000		&	50.14	&		&	84.30	&		&		&	1.7	&		\\
201pr1002		&	34.83	&	464.36	&	63.04	&		&	13.3	&	1.8	&		\\
207si1032		&	36.76	&	242.37	&	34.99	&		&	6.6	&	1.0	&		\\
212u1060		&	44.76	&	594.64	&	65.81	&		&	13.3	&	1.5	&		\\
217vm1084		&	59.82	&	562.04	&	87.38	&		&	9.4	&	1.5	&		\\
\hline																
Full IS total		&	321.0	&		&		&		&		&		&		\\
Sym. IS total/average		&	314.0	&		&	516.4	&		&		&	2.9	&		\\
SG IS total/average		&	237.1	&	2600.6	&	385.5	&		&	11.6	&	3.0	&		\\
TSP IS total/average		&	7.2	&	92.1	&	23.8	&	88.1	&	12.2	&	3.9	&	10.5	\\

\end{tabular}
\end{table}

The running times of the considered heuristics are presented in Table~\ref{tab:times}.  The running time of \texttt{GK} is not worse than the running time of any other heuristic for every instance: the minimum time improvement with respect to \texttt{SG} is 6.6 that is greater than 3 (recall that 3 is an adjusting coefficient for \texttt{SG} evaluation platform, see above), the time improvement with respect to \texttt{SD} is never less than 1.0 (recall that both heuristics were tested on the same platform), and the minimum time improvement with respect to \texttt{TSP} is 4.6 that is greater than 2 (recall that 2 is an adjusting coefficient for \texttt{TSP} evaluation platform, see above).  The time improvement average is $\sim$12~times for \texttt{SG} (or $\sim$4~times if we take into account the platforms difference), $\sim$3~times for \texttt{SD}, and $\sim$11~times for \texttt{TSP} (or $\sim$5~times if we take into account the platforms difference).

The stability of \texttt{GK} is high, e.g., for the \texttt{89pcb442} instance it produces only exact solutions and the time standard deviation is 0.27 sec for 100 runs.  The minimum running time is 1.29 sec, the maximum is 2.45 sec, and the average is 1.88 sec.  For 100 runs of \texttt{217vm1084} the average running time is 65.32 sec, the minimum is 44.30 sec, the maximum is 99.54 sec, and the standard deviation is 13.57 sec.  The average solution is 130994 (0.22\% above the best known), the minimum is 130704 (exactly the best known), the maximum is 131845 (0.87\% above best known), and the standard deviation is 331.

\begin{table}
	\caption{\texttt{GK} experiments details.}
	\label{tab:details}
\begin{tabular}{l|rrrrrr}

Name		&	Opt.	&	Value	&	Error, \%	&	Opt., \%	&	Time, sec	&	\# gen.	\\
\hline														

40d198		&	10557	&	10557.0	&	0.00	&	100	&	0.14	&	9.1	\\
40kroa200		&	13406	&	13406.0	&	0.00	&	100	&	0.14	&	9.0	\\
40krob200		&	13111	&	13111.0	&	0.00	&	100	&	0.16	&	10.3	\\
41gr202		&	23301	&	23301.0	&	0.00	&	100	&	0.21	&	9.8	\\
45ts225		&	68340	&	68340.0	&	0.00	&	100	&	0.24	&	12.7	\\
45tsp225		&	1612	&	1612.0	&	0.00	&	100	&	0.19	&	10.4	\\
46pr226		&	64007	&	64007.0	&	0.00	&	100	&	0.10	&	9.0	\\
46gr229		&	71972	&	71972.0	&	0.00	&	100	&	0.25	&	9.6	\\
53gil262		&	1013	&	1013.0	&	0.00	&	100	&	0.31	&	12.2	\\
53pr264		&	29549	&	29549.0	&	0.00	&	100	&	0.24	&	9.1	\\
56a280		&	1079	&	1079.0	&	0.00	&	100	&	0.38	&	13.1	\\
60pr299		&	22615	&	22615.0	&	0.00	&	100	&	0.42	&	11.9	\\
64lin318		&	20765	&	20765.0	&	0.00	&	100	&	0.45	&	12.8	\\
65rbg323	(asym.)	&	471	&	471.0	&	0.00	&	100	&	1.14	&	27.8	\\
72rbg358	(asym.)	&	693	&	693.0	&	0.00	&	100	&	1.26	&	24.4	\\
80rd400		&	6361	&	6361.0	&	0.00	&	100	&	1.07	&	15.0	\\
81rbg403	(asym.)	&	1170	&	1170.0	&	0.00	&	100	&	0.98	&	16.1	\\
84fl417		&	9651	&	9651.0	&	0.00	&	100	&	0.73	&	11.5	\\
87gr431		&	101946	&	101946.0	&	0.00	&	100	&	2.01	&	17.7	\\
88pr439		&	60099	&	60099.0	&	0.00	&	100	&	1.48	&	16.3	\\
89pcb442		&	21657	&	21657.0	&	0.00	&	100	&	1.72	&	21.2	\\
89rbg443	(asym.)	&	632	&	632.8	&	0.13	&	50	&	3.69	&	38.8	\\
99d493		&	20023	&	20044.8	&	0.11	&	10	&	4.17	&	27.3	\\
107ali535		&	128639	&	128639.0	&	0.00	&	100	&	5.82	&	25.1	\\
107att532		&	13464	&	13464.8	&	0.01	&	80	&	3.45	&	22.2	\\
107si535		&	13502	&	13502.0	&	0.00	&	100	&	1.88	&	19.5	\\
113pa561		&	1038	&	1038.0	&	0.00	&	100	&	3.22	&	22.2	\\
115u574		&	16689	&	16691.8	&	0.02	&	80	&	3.76	&	25.3	\\
115rat575		&	2388	&	2392.7	&	0.20	&	90	&	4.12	&	25.7	\\
131p654		&	27428	&	27428.0	&	0.00	&	100	&	2.82	&	15.3	\\
132d657		&	22498	&	22532.8	&	0.15	&	30	&	6.82	&	30.3	\\
134gr666		&	163028	&	163210.7	&	0.11	&	70	&	14.46	&	41.0	\\
145u724		&	17272	&	17296.8	&	0.14	&	50	&	11.61	&	38.9	\\
157rat783		&	3262	&	3265.7	&	0.11	&	20	&	15.30	&	40.1	\\
200dsj1000		&	9187884	&	9198846.6	&	0.12	&	30	&	50.14	&	49.1	\\
201pr1002		&	114311	&	114466.2	&	0.14	&	30	&	34.83	&	46.8	\\
207si1032		&	22306	&	22312.0	&	0.03	&	20	&	38.40	&	45.0	\\
212u1060		&	106007	&	106290.1	&	0.27	&	30	&	44.76	&	50.4	\\
217vm1084		&	130704	&	130954.2	&	0.19	&	60	&	59.82	&	50.5	\\
\hline														
Average		&		&		&	0.04	&	81	&		&	23.1	\\

\end{tabular}
\end{table}

Some details on the \texttt{GK} experiments are presented in Table~\ref{tab:details}.  The table includes the average number of generations produced by the heuristic.  One can see that the number of generations produced by our heuristic is relatively small: the \texttt{SD} and \texttt{TSP} limit the number of generation to 100 while they consider the instances with $M < 90$ only; \texttt{SG} terminates the algorithm after 150 idle generations.  Our heuristic does not require a lot of generations because of the powerful local search procedure and large population sizes.

\section{Conclusion}

We have developed a new memetic algorithm for GTSP that dominates all known GTSP heuristics with respect to both solution quality and the running time.  Unlike other memetic algorithms introduced in the literature, our heuristic is able to solve both symmetric and asymmetric instances of GTSP.  The improvement is achieved due to the powerful local search, well-fitted genetic operators and new efficient termination condition.

Our local search (LS) procedure consists of several LS heuristics of different power and type.  Due to their diversity, our algorithm is capable of successfully solving various instances.  Our LS heuristics are either known variations of GTSP heuristics from the literature (2-opt, Inserts, Cluster Optimization) or new ones inspired by the appropriate TSP heuristics (Swaps, $k$-Neighbor Swap, Direct 2-opt).  Note that our computational experiments demonstrated that the order in which LS heuristics are used is of importance.  Further research may find some better LS algorithms including more sophisticated based on, e.g., Tabu search or Simulated Annealing.

While crossover operator used in our algorithm is the same as in~\cite{GoldenGA}, the mutation operator is new.  The termination condition is also new.  The choices of the operators and the termination condition influence significantly the performance of the algorithm.

\bigskip
\noindent
\textbf{Acknowledgement}.
We would like to thank Natalio Krasnogor for numerous useful discussions of earlier versions of the paper and Michael Basarab for helpful advice on memetic algorithms.

\end{document}